\documentclass[%
 reprint,
 amsmath,amssymb,
 aps,
]{revtex4-1}
\usepackage{physics}
\usepackage{tikz}
\usepackage{float}
\usepackage{dsfont}
\usepackage{graphicx}% Include figure files
\usepackage{dcolumn}% Align table columns on decimal point
\usepackage{bm}% bold math
\usepackage{mathdots}
\newcommand{\ii}{\mathbf{i}}
\newcommand{\ee}{\mathbf{e}}
\begin{document}

\preprint{APS/123-QED}

\title{Construct Quantum Entanglement from Nonlocality}% Force line breaks with \\
%\thanks{Granted by...}%

\author{Ze Wu}
\email{wuze@mail.ustc.edu.cn}
\affiliation{Department of Modern Physics, University of Science and Technology of China, Hefei 230026, China}%Lines break automatically or can be forced with \\
\author{Ming-jun Shi}%
\email{shmj@ustc.edu.cn}
\affiliation{Department of Modern Physics, University of Science and Technology of China, Hefei 230026, China}%

\date{\today}% It is always \today, today,
             %  but any date may be explicitly specified

\begin{abstract}
``action at a distance'' is a weird concept in quantum theory which people always avoid mentioning in most occasions. In this work, without involving any concept about ``action at a distance'', we naturally construct a physical structure that resembles quantum entanglement only based on the nonlocality of the wave function. To achieve this, we propose a nonlocal structure of the single particle wave function by means of a double-well potential model. In order to better describe this nonlocal system, we have also developed two kinds of representations: local representation and nonlocal representation. Based on this nonlocal structure, we obtain a generalized exclusion principle in the case of antisymmetric nonlocal wave function of the multiparticle, with which we construct the entangled physical structure as mentioned before. In addition, we invent a diagram method to describe this entanglement structure and obtained some general conclusions.
\begin{description}
\item[PACS numbers]
03.65.Ta, 03.65.Ud

\end{description}
\end{abstract}

\pacs{Valid PACS appear here}% PACS, the Physics and Astronomy
                             % Classification Scheme.
%\keywords{Suggested keywords}%Use showkeys class option if keyword
                              %display desired
\maketitle

%\tableofcontents
\section{Introduction}
From EPR paradox\cite{RN1}, nonlocality in quantum machanics has been widely discussed. In most previous related studies, nonlocality usually refers to the inability to construct a local hidden variable theory reproducing the same prediction as quantum theory\cite{RN2-1,RN2-2}. However, the root cause of this inability is that the wave function separated over large distance can collapse instantaneously as soon as the measurement is implemented. In other words, the wave function itself is nonlocal whether it is of single particle or multiparticle\cite{RN3-1,RN3-2,RN3-3,RN3-4}. Moreover, previous opinion insists that the entangled particles are saparated in different places with some ``spooky action at a distance'' maintaining their entanglement\cite{RN4-1,RN4-2}. This is weird---not only about the ``action at a distance'', but, if their whole wave function is inseparable, why they are saparated in different places? In fact, there is an increasing number of questions about the rationality of the ``action at a distance''\cite{RN5-1,RN5-2}, and more discussion about the nonlocality and contextuality are taken in recent decades\cite{RN6-1,RN6-2,RN6-3,RN6-4}.

In our work, we hold that the nonlocality of the wave function allows each of these particles to locate in different places at the same time. Our main purpose is to construct a entangled theoretical structure only based on such a premise. First, through the discussion of the double-well potential, the intrinsic nonlocality of the single particle wave function is proposed. That is, the single particle wave function can be simultaneously defined at different locations which are far apart from each other. The property is also established in the multiparticle cases. And as we all know, the Pauli exclusion principle is a fundamental principle in quantum theory, which can be easily derived from the antisymmetry of the identical particles' wave function\cite{RN7-1,RN7-2}. This principle will be generalized if such an ``intrinsic nonlocality'' of the identical particles is taken into account. Based on this, a physical structure similar to the entangled state can be constructed without involving any concept related to the ``action at a distance''. 

\section{The intrinsic nonlocality of single particle wave function}
\begin{figure}[H]
	\centering
	\begin{tikzpicture}[>=stealth,scale=1.3,line width=1pt]
	\draw(-2,0)node[below]{$ -b $} -- (-0.75,0)node[below]{$ -a $} ;
	\draw(2,0)node[below]{$ b $} -- (0.75,0)node[below]{$ a $} ;
	\draw(-0.75,0) --(-0.75,1)--(0.75,1)--(0.75,0) ;
	\draw[->] (-2,0) -- (-2,1.5) node[left] {$+\infty$};
	\draw[->] (2,0)node[right]{$ -V_0 $} -- (2,1.5) node[right] {$ +\infty $};
	\draw[->,dashed] (0,0) -- (0,1.5) node[left] {$ V\left(x\right) $};
	\draw[->,dashed] (-2.2,1) -- (2.2,1) node[right] {$ 0 $};
	\draw[domain=-2:-0.75,smooth] plot (\x,{0.5*sin(400*(\x+2))});
	\draw[domain=0.75:2,smooth] plot (\x,{-0.5*sin(400*(\x-2))});
	\draw[domain=0:0.75,smooth,dashed] plot (\x,{0.01*exp(4.7*\x)});
	\draw[domain=-0.75:0,smooth,dashed] plot (\x,{0.01*exp(-4.7*\x)});
	\draw[domain=0.75:2,smooth,color=red] plot (\x,{0.5*sin(400*(\x-2))});
	\draw[domain=0:0.75,smooth,dashed] plot (\x,{-0.01*exp(4.7*\x)});
	\node at(0,-0.2){$ \psi\left(x\right) $};
	\node at(1.375,0.6){$ \text{Bob} $};
	\node at(-1.375,0.6){$ \text{Alice} $};
	\end{tikzpicture}
	\caption{Double-well potential model: The dashed part of $ \psi\left(x\right) $ decays rapidly when $ a $ increases. And the red part means the odd parity solution of this potential which differs with the other solution only by a minus sign (or $ \pi $ phase) in Bob's well, in the contex of $ a\rightarrow\infty $.}\label{fig-1}
\end{figure}
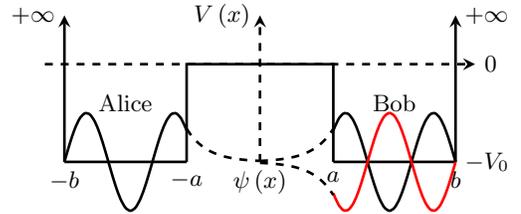

Let us look at a very simple physical model: double-well potential. The single particle wave function $ \psi\left(x\right) $ can be easily obtained by solving Schr\"{o}dinger equation. As Fig. \ref{fig-1} shows, $ \psi\left(x\right) $ is symmetrically distributed on both sides and its behavior is described by sine function in two wells ($a< \left| x\right| <b $) and by hyperbolic sine function between two wells ($ \left| x\right| <a $)\cite{RN8}.

Although there may exists the probability of detecting this particle in $ \left| x\right| <a $, the probability, however, is mostly distributed in or near the two wells. And as $ a $ increases, $ \left| \psi\left(x\right)\right|  $ will rapidly decay in the middle area. In fact, when $ a\rightarrow\infty $, the normalization coefficient will tend to be propotional to $ \ee^{-ka} $ ($ k>0 $), which is exactly the decay rate of $ \left| \psi\left(x\right)\right|  $ in the middle area. At this point we encountered a weird situation as Einstein said ``spooky actions at a distance'': when $ a $ increases into a large number, the particles can almost only be found in or near the area of two potential wells which can be at any distance from each other. Based on this fact, we have reason to claim that the single particle wave function can be simultaneously defined at different locations which are far apart from each other. We call this as the intrinsic nonlocality of single particle's wave function. Furthermore, as $ a\rightarrow\infty $, the nonlocal wave function's segment in each well tends to be propotional to the local wave function of single well (or semi-infinite well here exactly). Regardless of the phase freedom, they only differ by a $\frac{1}{\sqrt{2}}$ scaling factor. 

Now consider Alice and Bob each possess a potential well---Alice has the left one while Bob has the right one. Due to symmetry, the probability of finding the particle in their respective wells is $\frac{1}{2}$. Nobody knows which potential well the particle will appear in before any measurements are taken. However, once Alice has detected the particle, she can conclude that Bob will not detect the particle because the wave function has passed through the spacelike distance and collapsed into Alice's well, which also means Alice's behavior has influenced Bob's measurement results instantaneously.

This conclusion can be extended to $ n $-well model. A single particle's wave function can be evenly separated into $ n $ areas which are far apart from each other. The probability of detecting particle in or near each area is $\frac{1}{n}$. And more generally, this property is actually established in any kind of potential with $ n $ centers (n-center model).
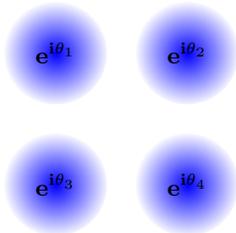
\begin{figure}[H]
	\centering
	\begin{tikzpicture}[>=stealth,scale=0.7,line width=1pt]
	\shadedraw[inner color=blue,outer color=white,draw=white] (0,0) circle (1cm);
	\shadedraw[inner color=blue,outer color=white,draw=white] (0,2.5) circle (1cm);
	\shadedraw[inner color=blue,outer color=white,draw=white] (2.5,0) circle (1cm);
	\shadedraw[inner color=blue,outer color=white,draw=white] (2.5,2.5) circle (1cm);
	\node at (0,0){$ \ee^{\ii\theta_3} $};
	\node at (0,2.5){$ \ee^{\ii\theta_1} $};
	\node at (2.5,0){$ \ee^{\ii\theta_4} $};
	\node at (2.5,2.5){$ \ee^{\ii\theta_2} $};
	\end{tikzpicture}
	\caption{4-well model: wave function segment in each well has a phase freedom.} \label{fig-2}
\end{figure}

Moreover, as the red part of Fig. \ref{fig-1} shows, the odd parity solution should not be ignored. When $ a $ is not very large, the even parity solution and the odd parity solution have different energy levels, which both change with $ a $. But as soon as $ a $ increases, these two sets of energy levels tend to be steady and merge together. In another word, two solutions will share the same energy levels as $ a\rightarrow\infty $. If we abandon mathematical rigor and think abous this fact in physical intuition, we can say the two solutions is a pair of degenerate states in the contex of $ a\rightarrow\infty $ which means these two kinds of wave functions only have symbolic (or phase) difference in half axis. Since we only care about these areas in or near the two wells, this difference can be regarded as a phase freedom of the wave function segment in each well. 

As for $ n $-well model or, more generally, $ n $-center model, although the Schr\"{o}dinger equation of it will encounter very complex boundary conditions, we have reason to believe that the nature of wave functions in double-well is still established here. On the one hand, the wave functions are evenly distributed in these potential wells; On the other hand, the nonlocal wave function's segment in each well can be regard as $ \frac{1}{\sqrt{n}} $ scaling down of the local wave function in such a single well excluding a phase factor as Fig. \ref{fig-2} shows.

\section{Local representation and nonlocal representation}
Until now, the physical model is clear, but there does not seem to be a corresponding mathematical expression. In order to describe the above physical model conveniently, we introduce two representation ways: local representation and nonlocal representation. And for convenience, all contents discussed below are in the context of $a\rightarrow\infty$, i.e., these potential wells are far apart from each other.

For $ n $-well model, we mark the areas of each well by $ 1,2\cdots,n $, the corresponding center coordinate of each well is $ \vec{x}_i $. For any $ i\neq j $, $ \left| \vec{x}_i-\vec{x}_j\right|\rightarrow\infty  $. The local wave function of well $ i $ is $ \ket{\psi_i} $, which means the particle described by $ \ket{\psi_i} $ can only be found near $ \vec{x}_i $. This representation is called as ``the local representation'' which is also the most common Dirac ket representation we widely used. However, ``Local'' does not mean that the wave function becomes localized. If we narrow our perspective inside the single potential well, the wave function is still nonlocal, but this time we are only concerned with the relationship between these potential wells which are far apart from each other instead of the details inside every single well, which, exactly can be equivalently regarded as an ideal model that is similar to a particle in dynamics.

Now, we mark $ n $-well's nonlocal wave function as $\ket{\psi^\prime} $. From the conclusion of section II and section III we know $ \ket{\psi^\prime} $ is nonlocal between these $ n $ potential wells, and the segment of $ \ket{\psi^\prime} $ in each well differs from the corresponding single well wave function $ \ket{\psi_i} $ only by a $ \frac{1}{\sqrt{n}} $ factor and a phase factor $ \ee^{\ii\theta_i} $:
\begin{equation*}
\text{Segment in well $ i $}=\frac{\ee^{\ii\theta_i}}{\sqrt{n}}\ket{\psi_i}
\end{equation*}
Since these potential wells are far apart, the value of $\ket{\psi_i}$ at well $j$ tends to be zero for any $i\neq j $, which makes $ \ket{\psi^\prime} $ can be expressed by the sum of its all segments.
\begin{equation}
\ket{\psi^\prime}=\sum_i\frac{\ee^{\ii\theta_i}}{\sqrt{n}}\ket{\psi_i}\label{3-1}
\end{equation}
We call $ \ket{\psi^\prime} $ as the nonlocal representation. The particle described by $ \ket{\psi^\prime} $ can possibly emerge in every wells which is the reason for being called ``nonlocal''.

This is exactly the same concept as the ``representation'' in Herbert space of quantum mechanics\cite{RN9}. Eq. \eqref{3-1} just represents the basis expansion, therefore the ready conclusions of the n-dimensional Hilbert space can be applied here. Firstly, $ \ket{\psi_i} $ is a set of basis vectors which satisfies orthogonality: 
\begin{equation*}
\braket{\psi_i}{\psi_j}=\delta_{ij}\mathds{1}
\end{equation*}
As Eq. \eqref{3-1} shows, all nonlocal wave function can be represented with these local bases' linear combination, and the phase freedom allows many possibilities of such combination. Hence we get another set of basis vectors: 
\begin{equation}
\ket{\psi^\prime_j}=\sum_{i}\frac{\ee^{\ii\theta_{ij}}}{\sqrt{n}}\ket{\psi_i}\label{3-2}
\end{equation}
where $ j=1,2,\cdots,n $, and the subscript of $ \ket{\psi^\prime} $ is no longer used to mark the potential well. Let $ C_{ij}=\frac{\ee^{\ii\theta_{ij}}}{\sqrt{n}} $ which is the elements of matrix $ C $, Eq. \eqref{3-2} can be rewritten in matrix form: 
\begin{equation}
\vec{\psi^\prime}=C\vec{\psi}\label{3-3}
\end{equation}
where $ \vec{\psi}=\left(\ket{\psi_1},\ket{\psi_2},\cdots,\ket{\psi_n}\right)^T $ is a column vector. In order to guarantee the orthogonality of $ \vec{\psi^\prime} $, $ C $ must be a unitary matrix, namely $ C^\dagger C =\mathds{1} $. Moreover, this Hilbert space is actually a degenerate space of a specific energy level, and each energy level has its corrosponding degenerate space and nonlocal basis vectors. 

Now that we have developed mathematical tools to represent nonlocal wave functions, until now, however, we have been talking about single particle. It is time to consider some multiparticle cases. In the following sections, we will see that the quantum entangled structure can be naturally constructed as long as we accept the nonlocality of the single particle wave function proposed in Section II. 

\section{The generalized exclusion principle}
Now consider two particles in a double-well. When $a$ is not large, the wave functions of the two particles must be evenly distributed in the two potential wells simultaneously. When $a$ increases, even tends to be infinite, there is no reason for this two particles' separating from each other and each occupying a potential well. This is naturally reminiscent of the nonlocal property, so can we use the nonlocal representation to represent two particles' wave functions?
\begin{figure}[H]
	\centering
	\includegraphics[width=0.35\textwidth]{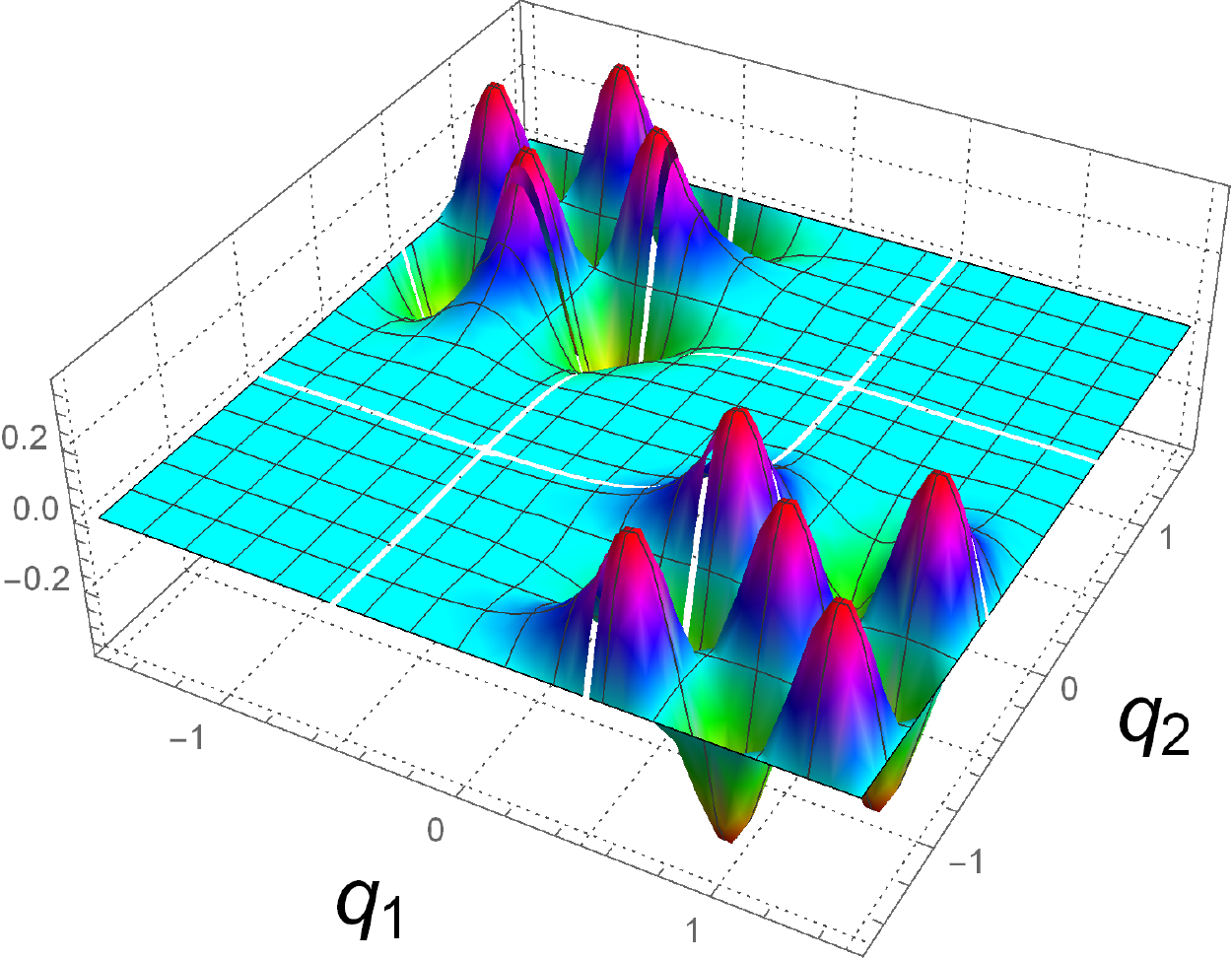}
	\caption{The $ \Psi^{\prime}\left(q_1,q_2\right) $ is only distributed in the domains of $ \left| q_1-q_2 \right|\rightarrow\infty  $, while it is canceled in the domains of $ q_1\approx q_2 $, which also means the two particles can't be found in the same potential well. This can be seen as a generalized exclusion principle, the two identical particles' mutual exclusion only happens in certain domains.} \label{fig-3}
\end{figure}

The direct product of the two nonlocal wave functions $\ket{\psi^\prime_1}\ket{\psi^\prime_2}$ is the simplest representation. Since the fundamental structure of matter is fermions, we will naturally consider the antisymmetric nonlocal wave function of two particles. Although Pauli exclusion principle forbids two identical fermions from being in the same quantum state, but they can be in different degenerate states and the nonlocal bases is just a set of degenerate states. Now suppose we have the following nonlocal basis vectors: 
\begin{equation}
\begin{cases}
\ket{\psi^\prime_1}&=\frac{1}{\sqrt{2}}\left(\ket{\psi_1}-\ket{\psi_2}\right)\\
\ket{\psi^\prime_2}&=\frac{1}{\sqrt{2}}\left(\ket{\psi_1}+\ket{\psi_2}\right)
\end{cases}\label{4-1}
\end{equation}
Their antisymmetry is: 
\begin{equation}
\begin{aligned}
\ket{\Psi^\prime}&=\frac{1}{\sqrt{2}}\left| \begin{pmatrix}
\mathds{1}\\ 
\mathds{1}
\end{pmatrix} 
\begin{pmatrix}
\ket{\psi^\prime_1} & \ket{\psi^\prime_2}
\end{pmatrix} \right| \\
&=\frac{1}{\sqrt{2}}\left(\ket{\psi^\prime_1}\ket{\psi^\prime_2}-\ket{\psi^\prime_2}\ket{\psi^\prime_1}\right)
\end{aligned}\label{4-2}
\end{equation}
Plug Eq. \eqref{4-1} in Eq. \eqref{4-2}, we found the terms of $ \ket{\psi_1}\ket{\psi_1} $ and $ \ket{\psi_2}\ket{\psi_2} $ are canceled, hence: 
\begin{equation}
\ket{\Psi^\prime}=\frac{1}{\sqrt{2}}\left(\ket{\psi_1}\ket{\psi_2}-\ket{\psi_2}\ket{\psi_1}\right)\equiv\ket{\Psi}\label{4-3}
\end{equation}
where $ \ket{\Psi} $ is the antisymmetry of local basis vectors. As we have agreed in Section III that the subscripts of the local basis vectors represent the potential well where they are located. The vanishing of $ \ket{\psi_1}\ket{\psi_1} $ and $ \ket{\psi_2}\ket{\psi_2} $ means these two states are forbidden here. In other words, two identical particles described by $ \ket{\Psi^\prime}$ cannot appear in the same potential well after measurement. The measurement result must be one in well 1 and the other in well 2. We label two particles with $q_1,q_2$, the state vector Eq. \eqref{4-3} can be rewritten in wave function form: 
\begin{equation*}
\Psi^{\prime}\left(q_1,q_2\right)=\Psi\left(q_1,q_2\right)=\frac{1}{\sqrt{2}}\left| \begin{matrix}
\psi_1\left(q_1\right)&\psi_2\left(q_1\right)  \\ 
\psi_1\left(q_2\right)& \psi_2\left(q_2\right)
\end{matrix}\right| 
\end{equation*}
This can be seen as a generalized exclusion principle as Fig. \ref{fig-3} shows. In addition, as an entirety, the two identical particles must collapse simultaneously. That is, if Alice finds a particle in well 1, she can conclude that Bob has also found another particle in well 2.
\begin{figure}[H]
	\centering
	\begin{tikzpicture}[>=stealth,scale=2,line width=1pt]
	\shade[ball color=blue](0,0) circle (.05);
	\shade[ball color=red](0,0.5) circle (.05);
	\shade[ball color=blue](1,0) circle (.05);
	\shade[ball color=red](1,0.5) circle (.05);
	\draw(0.05,0.5)--(0.95,0.5);
	\draw(0.05,0)--(0.95,0);
	\draw[dashed](0,0.05)--(0,0.45);
	\draw[dashed](1,0.05)--(1,0.45);
	\node at(0.5,0.08){$ q_1 $};
	\node at(0.5,0.58){$ q_2  $};
	\node at(0,0.7){$ 1 $};
	\node at(1,0.7){$ 2 $};
	\node at(-0.2,0){$ \left| 0\right\rangle  $};
	\node at(-0.2,0.5){$ \left| 1\right\rangle  $};
	%\node at(0.5,-0.2){$ \left| \Psi\right\rangle=\alpha\left| 10\right\rangle+\beta\left| 01\right\rangle $};
	\draw[->,line width=2pt](0.5,-0.1)--(0.5,-0.3);
	\node at(1,-0.2){Measuring};
	\begin{scope}[yshift=-1.2cm,xshift=-1cm]
	\shade[ball color=blue](0,0) circle (0.07);
	\shade(0,0.5) circle (0);
	\shade(1,0) circle (0);
	\shade[ball color=red](1,0.5) circle (0.07);
	\node at(0,0.7){$ 1 $};
	\node at(1,0.7){$ 2 $};
	\node at(-0.25,0){$ \left| 0\right\rangle  $};
	\node at(-0.25,0.5){$ \left| 1\right\rangle  $};
	\node at(0.5,-0.2){$ P_1=\dfrac{1}{2}$};
	\end{scope}
	\begin{scope}[yshift=-1.2cm,xshift=1cm]
	\shade(0,0) circle (0);
	\shade[ball color=red](0,0.5) circle (0.07);
	\shade[ball color=blue](1,0) circle (0.07);
	\shade(1,0.5) circle (0);
	\node at(0,0.7){$ 1 $};
	\node at(1,0.7){$ 2 $};
	\node at(0.5,-0.2){$ P_2=\dfrac{1}{2}$};
	\end{scope}
	\node at(0.5,-1.1){or};
	\end{tikzpicture}
	\caption{Two particles in 2-well model: Alice takes measurement in well 1 while Bob takes measurement in well 2. Blue balls represent the particle of ground state while red balls represent the one of excited state. For the generalized exclusion principle, the two particles cannot appear into the same well. There are two possible measurement results: (i). Alice in well 1 gets the ground state$ \ket{0} $ while Bob in well 2 gets the excited state$ \ket{1} $. (ii). Alice in well 1 gets the excited state$ \ket{1} $ while Bob in well 2 gets the ground state$ \ket{0} $. The probability of each event is $ \frac{1}{2} $.}\label{fig-4}
\end{figure}
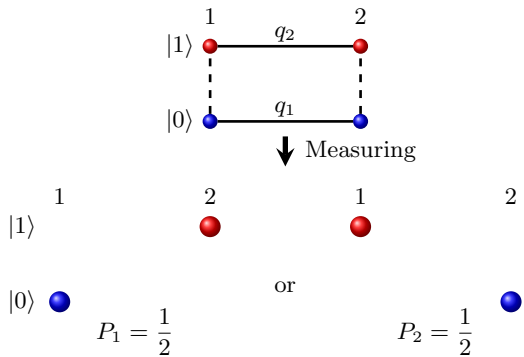

Now consider a more interesting situation: One of the particles is excited to other energy levels. We use $ \phi $ to represent the wave function of this energy level. Although we don't know exactly which particle was excited for the identicality, we can assume for convenience that the particle labeled with $ q_2 $ is excited, and any conclusions is the same with the identical substitution. In this way, is the above nature still established? The answer is positive. Let the wave function under this circumstance be $\ket{\Omega^\prime} $. In Eq. \eqref{4-2}, we just substitute the state vector to the right of the direct product (or the $ q_2 $'s direct product space) with $ \ket{\phi^\prime_i} $, we can see: 
\begin{equation*}
\begin{aligned}
\ket{\Omega^\prime}&=\frac{1}{\sqrt{2}}\left(\ket{\psi^\prime_1}\ket{\phi^\prime_2}-\ket{\psi^\prime_2}\ket{\phi^\prime_1}\right)\\
&=\frac{1}{\sqrt{2}}\left(\ket{\psi_1}\ket{\phi_2}-\ket{\psi_2}\ket{\phi_1}\right)
\end{aligned}
\end{equation*}
Similarly, the terms of $ \ket{\psi_1}\ket{\phi_1} $ and $ \ket{\psi_2}\ket{\phi_2} $ are canceled, which, as before, means that two particles cannot appear in the same potential well after measurement. The wave function at this time can be expressed as: 
\begin{equation*}
\Omega^\prime\left(q_1,q_2\right)=\frac{1}{\sqrt{2}}\left| \begin{matrix}
\psi_1\left(q_1\right)&\psi_2\left(q_1\right)  \\ 
\phi_1\left(q_2\right)& \phi_2\left(q_2\right)
\end{matrix}\right| 
\end{equation*}
Therefore, the measurement of $ \Omega^\prime\left(q_1,q_2\right)$ can only get two results: (i). The particle of $ \phi $ appears in well 1 while the particle of $ \psi $ appears in well 2 ; (ii). the particle of $ \phi $ appears in well 2 while the particle of $ \psi $ appears in well 1.

As shown in Fig. \ref{fig-4}. If $\ket{\psi}$ is taken as $\ket{0} $ and $\ket{\phi}$ is taken as $\ket{1} $, then the measurement property of $\ket{\Omega^\prime} $ becomes very similar to the two qubits' entangled state $ \frac{1}{\sqrt{2}}\left(\ket{01}+\ket{10}\right)$. In this theoretical structure, some concepts are different with the previous consensus. Before the measurement, none of Alice and Bob can even possess a integral particle. They are actually faced up with the two wave function's segments of each particle. But after the measurement they can only get one particle, and which one (0 or 1) to get is random.

\section{Generalizations}
With the help of nonlocal representation, the conclusions above can be easily generalized to $ n $ particles in $ n $-well model. Consider about the transposed form of Eq. \eqref{3-3}:
\begin{equation*}
	\left(\ket{\psi^\prime_1},\ket{\psi^\prime_2},\cdots,\ket{\psi^\prime_n}\right)=\left(\ket{\psi_1},\ket{\psi_2},\cdots,\ket{\psi_n}\right)C^T
\end{equation*}
Where $ C^\dagger C =\mathds{1} $. With this Eq. \eqref{4-2} is generalized as follows:
\begin{equation*}
	\ket{\Psi^{\prime(n)}}=\frac{1}{\sqrt{n!}}\left|
	\begin{pmatrix}
		\mathds{1}\\ 
		\mathds{1}\\ 
		\vdots\\ 
		\mathds{1}
	\end{pmatrix} 
	\begin{pmatrix}
		\ket{\psi_1}&\ket{\psi_2}&\cdots&\ket{\psi_n}
	\end{pmatrix} 
	C^T
	\right|
\end{equation*}
Utilizing the nature of the determinant $ \left|AB\right|=\left|A\right|\left|B\right| $ and $ \left| C^{T}\right|=\left| C\right|   $, we have:
\begin{equation*}
\ket{\Psi^{\prime(n)}}=\ket{\Psi^{(n)}}\left| C\right|
\end{equation*}
where $ \ket{\Psi^{(n)}} $ is the antisymmetry of the local basis vectors. As we already have $ CC^\dagger=\mathds{1} $ which derives $ \left|C\right|=\ee^{\ii \Theta} $. This term only contributes an overall phase factor and can be ignored here. Thus we still have $ \ket{\Psi^{\prime(n)}}=\ket{\Psi^{(n)}} $ which means the generalized exclusion principle is still established here. The measurement result of this system is that each potential well can only get one of these identical particles. Similarly, by exciting one of the particles to other levels, we can get some kind of entangled structures. And as Section V does, although we don't know which particle was excited, symmetry allows us to pick one of them arbitrarily which, mathematically, is to change one of the unit operators of $ \begin{pmatrix}
\mathds{1}&\mathds{1}&\cdots&\mathds{1}
\end{pmatrix}^T $ into another operator $ \hat{O} $, which makes:
\begin{equation}
	\begin{aligned}
		\ket{\Omega^{\prime(n)}}&=\frac{\left| C\right|}{\sqrt{n!}}\left|
		\begin{pmatrix}
			\hat{O}\\ 
			\mathds{1}\\ 
			\vdots\\ 
			\mathds{1}
		\end{pmatrix} 
		\begin{pmatrix}
			\ket{\psi_1}&\ket{\psi_2}&\cdots&\ket{\psi_n}
		\end{pmatrix}
		\right|\\
		&\equiv \ee^{\ii\Theta}\ket{\Omega^{(n)}}
	\end{aligned}\label{5-1}
\end{equation}
Now we suppose that $\ket{\psi}$ corresponds to $\ket{0} $ and $\hat{O}$ represents a NOT gate\cite{RN10}. Under the circumstances we get a $n$-particle state whose behavior is very similar to the $ W $ state for n qubits\cite{RN11-1,RN11-2}. 
\begin{figure}[H]
	\centering
	\begin{tikzpicture}[>=stealth,scale=2]
	\shade[ball color=blue](0,0) circle (.05);
	\shade[ball color=red](0,0.5) circle (.05);
	\shade[ball color=blue](0,-0.1) circle (.05);
	
	\shade[ball color=blue](1,0) circle (.05);
	\shade[ball color=red](1,0.5) circle (.05);
	\shade[ball color=blue](1,-0.1) circle (.05);
	
	\shade[ball color=blue](0.6,0.3) circle (.05);
	\shade[ball color=red](0.6,0.8) circle (.05);
	\shade[ball color=blue](0.6,0.2) circle (.05);
	
	\draw(0,0.5)--(1,0.5)--(0.6,0.8)--(0,0.5);
	\draw(0,0)--(1,0)--(0.6,0.3)--(0,0);
	\draw(0,-0.1)--(1,-0.1)--(0.6,0.2)--(0,-0.1);
	\draw[dashed](0,0)--(0,0.5);
	\draw[dashed](1,0)--(1,0.5);
	\draw[dashed](0.6,0.3)--(0.6,0.8);
	
	\node at(0,0.65){$ 1 $};
	\node at(1,0.65){$ 2 $};
	\node at(0.6,0.95){$ 3 $};
	
	\node at(-0.2,-0.05){$ \left| 0\right\rangle  $};
	%	\node at(-0.2,-0.1){$ \left| 0\right\rangle  $};
	\node at(-0.2,0.5){$ \left| 1\right\rangle  $};
	
	\draw[->,line width=2pt](1.2,0.25)--(1.8,0.25);
	\node at(1.5,0.35){Measuring};
	
	\begin{scope}[xshift=2cm]
	\shade[ball color=blue](0,-0.1) circle (.07);
	
	\shade[ball color=blue](1,0) circle (.05);
	\shade[ball color=red](1,0.5) circle (.05);
	
	\shade[ball color=blue](0.6,0.3) circle (.05);
	\shade[ball color=red](0.6,0.8) circle (.05);
	
	\draw(1,0.5)--(0.6,0.8);
	\draw(1,0)--(0.6,0.3);
	\draw[dashed](1,0)--(1,0.5);
	\draw[dashed](0.6,0.3)--(0.6,0.8);
	
	\node at(0,0.65){$ 1 $};
	\node at(1,0.65){$ 2 $};
	\node at(0.6,0.95){$ 3 $};
	
	\end{scope}
	\end{tikzpicture}
	\caption{Three particles in 3-well model: The particle $ q_{1} $ was excited to $ \ket{1} $. After Alice took measurement in well 1, supposing $ q_3 $ appeared in well 1 which means Alice has got $ \ket{0} $. Then Bob in well 2 and Cindy in well 3 will be faced up with a two particle entangled state as Fig. \ref{fig-4} shows.}\label{fig-5}
\end{figure}
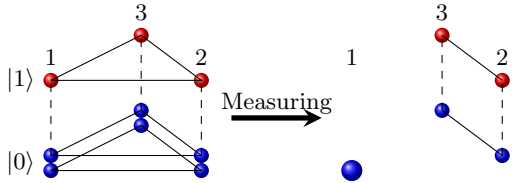
Let us take $ n=3 $ for example. It's nonlocl wave function is: 
\begin{equation*}
		\ket{\Omega^{\prime(3)}}=\frac{1}{\sqrt{6}}\left| \begin{matrix}
			\ket{1_{1}}& \ket{1_{2}} & \ket{1_{3}} \\ 
			\ket{0_{1}}& \ket{0_{2}} & \ket{0_{3}} \\ 
			\ket{0_{1}}&\ket{0_{2}}  & \ket{0_{3}}
		\end{matrix} \right|
\end{equation*}
As before, the subscript here is used to mark different potential wells. Using algebraic cofactor to calculate the determinant will make our understanding of this state more clear and organized. As Fig. \ref{fig-5} shows, Alice in well 1 got $ \ket{0_{1}} $ whose algebraic cofactor is $ \left(\ket{1_{2}}\ket{0_{3}}-\ket{1_{3}}\ket{0_{2}}\right) $, which is the very state faced by Bob in well 2 and Cindy in well 3 and it's exactly the state of Fig. \ref{fig-4}. If Alice got $\ket{1_{1}}$, then Bob in well 2 and Cindy in well 3 would be facing $\left(\ket{0_2}\ket{0_3}- \ket{0_3}\ket{0_2}\right) $, which can be regarded as an entangled structure at the same energy level. Since $\ket{0_2}\ket{0_3}$ and $\ket{0_3}\ket{0_2} $ cannot be distinguished after measurement, their measurement result is the same as $\ket{00}$. In summary, the behavior of $ \ket{\Omega^{\prime(3)}} $ and $ W $ state are alike. If viewed in the sense of measurement, they can actually represent the same type of entangled state.

Now if we excite more particles to $ \ket{1} $ in general cases, a variety of entangled structure will be constructed. In the next Section we will briefly discuss the relationship of these entangled states.
\section{Entanglement diagrams and Pascal's triangle}
At this point, the core object of our research is a matrix determinant like Eq. \eqref{5-1}. For the general case, we can also use some diagrams to illustrate their measurement properties visually as Fig. \ref{fig-3} and Fig. \ref{fig-4} did. The nonlocal wave function of each particle in $ n $-well corresponds to a complete graph of $ n $ vertices. Each vertex represents the wave function segment in each potential well. When measurement is taken in one of these wells, there must be a particle collapse into this well, and the remaining particles are no longer likely to appear in this well. Such measurement makes all of the remaining particles become complete graphs of $ n-1 $ vertices, and the $ n-1 $ vertices are located respectively in the remaining $ n-1 $ potential wells. Judging from the determinant, the state of the remaining $ n-1 $ particles is exactly the algebraic cofactor of the already measured state. The case of $ n $ and $ n-1 $ also constitute a recurrence relation which has the following Pascal triangle structure:
\begin{figure}[H]
	\centering
\begin{tabular}{*{9}{c}}
	&&&$ \ket{0} $&&$ \ket{1} $&&&\\
	&&$ \ket{00} $&& $ B $ &&$ \ket{11} $&&\\
	&$ \ket{000} $&&$ W_{+}^{(3)} $&&$ W_{-}^{(3)} $&&$ \ket{111} $&\\
	$ \ket{0000} $&&$ W_{+}^{(4)} $&&$ Z $&&$ W_{-}^{(4)} $&&$ \ket{1111} $\\
	$ \iddots $&&&&$ \vdots $&&&&$ \ddots $
\end{tabular}
\caption{Pascal's triangle of the entangled structure: Excluding the normalization coefficients, B means one of the Bell states: $ \ket{10}+\ket{01} $; $ W_{+}^{(n)} $ means the $ W $ state for $ n $ qubits with only one $ \ket{1} $ and $ W_{+}^{(n)} $ means that with only one $ \ket{0} $.}\label{fig-6}
\end{figure}
In Fig. \ref{fig-6}, $ n $ gradually increases from the top down; and from left to right, the number of excited particles (or the number of $ \ket{1} $) is gradually increased. Each of these states can be represented by the two states on its shoulder:
\begin{equation*}
	\ket{\text{Each state}}=\ket{1}\ket{\text{Left shoulder}}+\ket{0}\ket{\text{Right shoulder}}
\end{equation*}
When $n=4 $, there will be a rarely-discussed four-bit entangled state in addition to the $ W $ states. We call it the $ Z $ state: 
\begin{equation*}
	\begin{aligned}
		\ket{Z}&=\frac{1}{\sqrt{6}}\left(\ket{1}W_{+}^{(3)}+\ket{0}W_{-}^{(3)}\right)\\
		&=\frac{1}{\sqrt{6}}\left( \ket{1100}+\ket{1010}+\ket{1001}\right. \\
		&\qquad\qquad \left. +\ket{0011}+\ket{0101}+\ket{0110}\right) 
	\end{aligned}
\end{equation*}
After measuring any one of the four qubits, the remaining three qubits are bound to be in the $W $ state. Someone has already studied this state before, and concluded that it is the maximally entangled state of four qubits\cite{RN12-1,RN12-2}.

As the Pascal triangle is written down along with $ n $, more and more entangled states will appear, which has a very good guiding significance for the study of this entanglement structure. However, some readers may wonder why the quantum state like $\ket{00}+\ket{11}$ is not inside. If we carefully review our construction process, we will find that all these states are constructed in the degenerate space. $\ket{01}$ and $\ket{10}$ are a couple of degenerate states, while $\ket{00}$ and $\ket{11}$ are not.
\section{Conclusion and expectation}
Until now, our wording is very conservative, and we have always regarded this work as a mathematical construct without any ambitious attempt to overturn previous concepts. But if we look at this mathematical process from the perspective of physics, can we question the rationality of the previous description of quantum entanglement that the two entangled particles each occupy two places that are far apart and then get entangled by some kind of ``action at a distance''? Or is quantum entanglement merely a manifestation of the nonlocality of identical particles' wave function? Of course, this work alone cannot answer these questions. We need more theoretical research, such as discussing the entanglement degree, the entropy, and Bell inequalities under this structure; In experiment, ion traps may also helpful to bulid this entangled structure in lab\cite{RN13}. In any case, we hope that our work will open up another perspective on quantum entanglement which is more natural than the ``action at a distance'' of the previous views.
\bibliography{apssamp}% Produces the bibliography via BibTeX.

\end{document}